\documentclass[fleqn,10pt,onecolumn]{wlscirep}

\usepackage[utf8]{inputenc}
\usepackage[T1]{fontenc}
\usepackage[acronym]{glossaries}
\usepackage{physics}
\usepackage{caption}
\usepackage{subcaption}

\glsdisablehyper
\newacronym{dft}{DFT}{density-functional theory}
\newacronym{dos}{DOS}{density of states}
\newacronym{sorep}{SOREP}{spectral operator representation}
\newacronym{mc3d}{MC3D}{Materials Cloud 3D database}
\newacronym{tcm}{TCM}{transparent conducting material}
\newacronym{cnt}{CNT}{carbon nanotube}
\newacronym{ml}{ML}{machine learning}
\newacronym{spahm}{SPA$^\mathrm{H}$M}{spectrum of approximated Hamiltonian matrices}
\newacronym{soap}{SOAP}{smooth overlap of atomic positions}
\newacronym{om}{OM}{overlap matrix}
\newacronym{qe}{QE}{QuantumESPRESSO}
\newacronym{ks}{KS}{Kohn-Sham}
\newacronym{hf}{HF}{Hartree-Fock}
\newacronym{lcao}{LCAO}{linear combination of atomic orbitals}
\newacronym{gto}{GTO}{Gaussian-type orbital}
\newacronym{cgto}{cGTO}{contracted Gaussian-type orbital}
\newacronym{pao}{PAO}{pseudo-atomic orbital}
\newacronym{ano}{ANO}{atomic natural orbital}
\newacronym{dbscan}{DBSCAN}{density-based spatial clustering of applications with noise}
\newacronym{sssp}{SSSP}{standard solid-state pseudopotential}
\newacronym{vbm}{VBM}{valence band maximum}
\newacronym{cbm}{CBM}{conduction band minimum}
\newacronym{homo}{HOMO}{highest occupied molecular orbital}
\newacronym{lumo}{LUMO}{lowest unoccupied molecular orbital}
\newacronym{knn}{$k$-NN}{$k$-nearest neighbors}
\newacronym{scf}{SCF}{self-consistent field}
\newacronym{rfc}{RFC}{random forest classifier}
\newacronym{ke}{KE}{kinetic energy}
\newacronym{gga}{GGA}{generalized gradient approximation}

\newcommand\cfref[1]{Fig.~\ref{#1}}

\newcommand\ctref[1]{Tab.~\ref{#1}}

\newcommand\ceref[1]{Eq.~\ref{#1}}
\newcommand\crref[1]{Ref.~\citeonline{#1}}

\newcommand{\etal}{\textit{et al}.}

\newcommand{\editor}[2]{
  \expandafter\newcommand\csname #1note\endcsname[1]{%
    \textcolor{#2}{(\textbf{#1:} ##1)}}%
  \expandafter\newcommand\csname #1\endcsname[1]{%
    \textcolor{#2}{##1}}%
  \expandafter\newcommand\csname #1cancel\endcsname[1]{%
    \textcolor{#2}{\sout{##1}}}%
  \expandafter\newcommand\csname #1change\endcsname[2]{%
    \textcolor{#2}{\sout{##1} ##2}}%
  \newenvironment{#1text}{\color{#2}}{\color{black}}
}
\editor{AZ}{blue}
\editor{NM}{red}
\editor{AM}{teal}

\title{Spectral Operator Representations}

\author[1]{Austin Zadoks}
\author[2]{Antimo Marrazzo}
\author[1,3,4]{Nicola Marzari}

\affil[1]{Theory and Simulation of Materials (THEOS), \'Ecole Polytechnique F\'ed\'erale de Lausanne, CH-1015 Lausanne, Switzerland}
\affil[2]{Dipartimento di Fisica, Universit\`a di Trieste, I-34151 Trieste, Italy}
\affil[3]{National Centre for Computational Design and Discovery of Novel Materials (MARVEL), \'Ecole Polytechnique F\'ed\'erale de Lausanne, CH-1015 Lausanne, Switzerland}
\affil[4]{Laboratory for Materials Simulations (LMS), Paul Scherrer Institut, CH-5232 Villigen, Switzerland}

\begin{abstract}
    Machine learning in atomistic materials science has grown to become a powerful tool, with most approaches focusing on atomic arrangements, typically decomposed into local atomic environments.
    This approach, while well-suited for machine-learned interatomic potentials, is conceptually at odds with learning complex intrinsic properties of materials, often driven by spectral properties commonly represented in reciprocal space (e.g., band gaps or mobilities) which cannot be readily atomically partitioned.
    For such applications, methods which represent the electronic rather than the atomic structure could be more promising.
    In this work, we present a general framework focused on electronic-structure descriptors which take advantage of the natural symmetries and inherent interpretability of physical models.
    Using this framework, we formulate two such representations and apply them respectively to measuring the similarity of carbon nanotubes and barium titanate polymorphs, and to the discovery of novel transparent conducting materials (TCMs) in the Materials Cloud 3D database (MC3D).
    A random forest classifier trained on 1\% of the materials in the MC3D is able to correctly label 76\% of entries in database which meet common screening criteria for promising TCMs.
\end{abstract}

\begin{document}

\flushbottom
\maketitle

\section*{Introduction}
The past two decades have seen an explosion in the amount and availability of databases materials structures properties database \cite{Jain_2013, Hellenbrandt_2004, Wishart_2008, Grazulis_2009, Curtarolo_2012, Saal_2013, Kim_2016, Puchala_2016, Borysov_2017, Villars_2018, Mendez_2018, Draxl_2019, Choudhary_2020, Talirz_2020}.
Simultaneously, the infrastructure and protocols for performing high-throughput studies have matured and now allow for producing large volumes of high-quality data with ease \cite{ASE_2002, ASE_2017, Pizzi_2016, Huber_2020, Uhrin_2021, Jain_2015, Kiran_2017, Huber_2021}.
As in computer vision, natural language processing, and other fields where the combination of data availability and \gls{ml} techniques have enabled powerful technologies from autonomous driving to machine translation, data-driven materials science is a promising new approach for accelerated materials discovery, property prediction, and inverse design.

This promise is somewhat tempered by the unique problem of representation: traditional ``xyz'' descriptions of atomic configurations as a set of atomic positions and species $\{\mathbf{R}_I, \alpha_I\}$ cannot be used directly to efficiently drive traditional statistical models.
Properties of atomic systems, like total energy and forces are either invariant or equivariant to rotations and translations of atoms and to permutations in the order in which they are listed, while the atomic coordinates and species are not.
Therefore, one of the foremost problems in data-driven materials science is how to efficiently and compactly describe relevant information of an atomic system in a framework suited to \gls{ml} applications.
There exist a few broad categories of approaches to solving this problem \cite{Musil_2021}, including atomic-density field features like the \gls{soap} \cite{Bartok_2013}, internal coordinates representations like Behler-Parrinello symmetry functions \cite{Behler_2007,Behler_2011}, atomic cluster expansions \cite{Drautz_2019}, and end-to-end neural network models, often based on atom graphs, like the CGNN \cite{Xie_2018}, NequIP \cite{Musaelian_2023}, or MACE \cite{Batatia_2022} models.
All of these approaches take atomic structures as the fundamental objects to process into inputs for \gls{ml} models, and most decompose them into atom-centered motifs for the purpose of imposing translational invariance and for aiding transferability.
These approximations are well-founded for learning problems where the target property or properties are extensive or conceptually decomposable into atomic contributions.
However, these approaches can be limited by their strong scaling with compositional complexity, degeneracies in the local atomic expansion at low body-orders \cite{Pozdnyakov_2020}, and by the fundamental concept of atomic decomposition, which struggles to capture important electronic quantities like band gap, quasi-particle energies, electrical conductivity, or optical properties, to name a few.

For these applications, a class of descriptors that can capture the physics and interactions pertaining to these complex properties would be beneficial.
A promising approach can be seen in methods which leverage electronic structure for featurization.
Methods in this class include the \gls{spahm} \cite{Fabrizio_2021,Fabrizio_2022}, the D-fingerprint \cite{Isayev_2015}, and moments of the density of states \cite{Hammerschmidt_2016, Jenke_2018}, among many others \cite{Fung_2021,Knosgaard_2022,Geilhufe_2018}, with successful applications to structure similarity\cite{Geilhufe_2018,Isayev_2015}, regression of various quantum-chemical properties \cite{Fabrizio_2022}, and delta learning of G$_0$W$_0$ quasi-particle energies \cite{Knosgaard_2022}.
These algorithms follow the general approach of computing the spectrum of a physical operator applied to a model electronic structure followed by a transformation into a machine learning descriptor.
We formalize and generalize this process in a framework for designing electronic-structure features, which we call \glspl{sorep}.

\section*{Results}
\label{sec:results}

\subsection*{SOREP Framework}
\label{sec:framework}

\glspl{sorep} aim to describe materials using targeted features of their electronic structure.
However, neither experimental nor high-throughput \textit{ab-initio} materials databases generally provide detailed electronic structure objects (e.g. Kohn-Sham orbitals), so the process begins with knowledge of only atomic structure.
The first step in featurizing a material with a \gls{sorep} is to build a model electronic structure from atomic positions.
The quality of this electronic-structure calculation determines the quality of the information content of the representation -- a consideration that must be carefully balanced in terms of the corresponding computational cost of directly determining the quantity of interest.
Predicting more complex physical phenomena may require more expensive, but more faithful, electronic-structure representations, while applications involving millions of systems might necessitate more cost-effective approximations.
In general, this first step entails applying some map $f$ of the atomic structure (positions, species, etc.) which yields, in principle, a many-body wavefunction or another representation of the electronic structure (e.g. a density matrix or Green's function)
\begin{equation}
    f: \{\mathbf{R}_I, \alpha_I, ...\} \rightarrow \Psi(\mathbf{r}_1, \mathbf{r}_2, \dots, \mathbf{r}_N).
\end{equation}
However, for \gls{ml} applications, many-body electronic-structure calculations are impractical.
A more pragmatic approach, and the one we will consider in moving forward, is to generate set of single-particle orbitals from the atomic structure:
\begin{equation}
    f: \{\mathbf{R}_I, \alpha_I, ...\} \rightarrow \{\ket{\phi_i}\}.
\end{equation}
This electronic system, however it may be represented, exists in a much higher-dimension space than its originating atomic configuration (an atomic structure of $N$ atoms can be considered to exist in a $3N$-dimensional Cartesian space, while in principle its electronic wavefunction exists in the Hilbert space of the problem considered).
In order to extract compact and useful information from this raw electronic structure, a Hermitian operator $\hat{A}$ selected from physical intuition or constructed through careful engineering can be projected onto the set of orbitals
to compute the operator matrix elements
\begin{equation}
    A_{ij} = \matrixel{\phi_i}{\hat{A}}{\phi_{j}}.
\end{equation}
$\hat{A}$ may be simple and efficient to evaluate, like the identity or kinetic energy operators, more expressive yet expensive like the Kohn-Sham Hamiltonian, or somewhere in between like the various guess Hamiltonians explored in \crref{Fabrizio_2022}.
Here, we consider only scalar operators (in the physical sense, i.e. independent of changes in frame of reference) in order to achieve rotation and translation invariance of the matrix elements.
A further generalization can be made to higher-order tensor operators, like position, if the features are to be used in an equivariant model or if further consideration is taken in the following steps to enforce reference-frame invariance.
Regardless, the resulting operator matrix $A$ represents a distillation of the electronic structure, filtered through the lens of the operator and expressed in the chosen basis.

To make use of all the information contained within the operator matrix, one could consider leveraging the matrix elements $A_{ij}$ as \gls{ml} features, as explored in \crref{Knosgaard_2022}.
Although invariant to translation, rotation, and other physically relevant symmetries (because $\hat{A}$ is scalar), the matrix elements are sensitive the choice of the basis functions $\ket{\phi_i}$.
So, a key step in formulating a \gls{sorep}, mirroring the standard procedure for electronic-structure calculations, is to diagonalize the operator matrix
\begin{equation}
    A \ket{\phi_i} = \lambda_i S \ket{\phi_i},
\end{equation}
using the overlap matrix $S$, to retrieve its set of eigenvalues $\{\lambda_i\}$.
This procedure removes explicit dependence on the choice of basis (for complete bases), and, significantly, it also mixes the information contained in the operator matrix in a non-trivial and physically meaningful manner \cite{Sadeghi_2013,Zhu_2016}.
In order to bring the eigenspectrum into a system-independent constant-dimensional space, as is required by all ML models, and to enforce invariance to permutations of the eigenvalue indices, the final step of the \gls{sorep} procedure is to apply a map $g$ from the set of eigenvalues $\lambda_i$ to a feature vector $\mathbf{x}$:
\begin{equation}
    g: \{\lambda_i\} \rightarrow \mathbf{x}.
\end{equation}
One simple and compact method for systems with few eigenvalues is to sort the spectrum and pad it with zeros up to a common constant dimension, as done in the \gls{spahm} method.
However, the resulting features are discontinuous w.r.t. level crossings and are high-dimensional for systems where many eigenstates are considered (e.g. periodic systems sampled at many k-points and/or with many bands).
To remedy these shortcomings, a \gls{dos} computed on a basis, e.g. as a sum-over-poles \cite{Chiarotti_2022} or using polynomials \cite{Musil_2021}, and sampled on a fixed domain can be more compact and is smooth w.r.t. level crossings.
Other maps used in the literature are spectral histograms \cite{Kuban_2022,Isayev_2015,Fung_2021,Geilhufe_2018}, moments of the density of states \cite{Sadeghi_2013,Hammerschmidt_2016}, and radially-decomposed projected densities of states \cite{Knosgaard_2022}.

In general, \glspl{sorep} exhibit many desirable properties for atomistic descriptors ``for free'' due to the properties of spectra \cite{Musil_2021}.
Key symmetry invariances, such as to rigid translation and rotation, are ensured by construction through utilization of scalar operators.
Beyond respecting physical symmetries, atomic descriptors should be complete; i.e. they should always distinguish (symmetry-)inequivalent structures.
It has been shown \cite{Pozdnyakov_2020} that low body-order local atomic descriptors can suffer to an extent from incompleteness, mapping distinct configurations to (nearly) identical descriptors.
\Gls{om} fingerprints, as atom-centered spectral representations which leverage the overlap (identity) operator, have been seen in practice to lift these degeneracies \cite{Musil_2021,Parsaeifard_2022}.
However, the limits of the completeness of spectral representations, in particular of global (i.e. not atomically-decomposed) fingerprints, have not been rigorously bounded \cite{Pozdnyakov_2022b}.
Additionally, because many properties of interest (e.g. total energies) vary smoothly with continuous deformations of the atomic structure, feature maps are often constructed to be similarly smooth.
It is key to note that this criterion is intended to ensure that no \textit{nonphysical} discontinuities can be found in the feature map, which could lead to, e.g., spurious discontinuities in a learned potential energy surface.
Some electronic properties, like band gaps and Van Hove singularities, may not be smooth w.r.t. structural deformations.
Unlike local atomic descriptors, spectral representations can capture these physical discontinuities and should be better suited for learning similarly nonsmooth properties.

To provide a more direct understanding of what this procedure entails and to show how specific constraints influence choices in each of the steps, we next consider the case where one aims to minimize as much as possible the computational cost of featurization while maintaining a spectral representation.

\subsection*{Kinetic energy SOREP}
Often, the featurization of millions of structures may be required in order to apply ML to a given problem, for example in testing structure uniqueness or learning from frames of molecular dynamics trajectories.
In these situations low-cost featurizations are essential, so here we discuss how to design a \gls{sorep} with this constraint in mind.
Generally, the diagonalization of operator matrices is the most computationally demanding step in producing \gls{sorep} features, but, as discussed above, it is essential in enforcing various symmetry invariances and in capturing non-local properties.
Therefore, the important ingredients to consider for optimization are the choices of how to determine and represent the electronic structure, which operator to apply, and how to map the operator spectrum onto features after diagonalization.

An appealing electronic-structure model for these purposes is a linear combination of \glspl{cgto}, for which many basis sets have been constructed alongside efficient libraries like \texttt{libcint} \cite{Sun_2015} for applying operators and computing integrals analytically.
A \gls{cgto} with quantum numbers $n,l,m$ for atom of species $\alpha$ is constructed as the product of a spherical harmonic $Y_l^m(\theta,\phi)$ and a radial function
\begin{equation}
    R_{nl}^{\alpha}(r) = r^l~\sum_{p} c_p^{\alpha}~B(l, a_p^{\alpha})~e^{-a_p^{\alpha} r^2}
\end{equation}
where $c_p^\alpha$ and $a_p^\alpha$ are the contraction coefficient and exponent for species $\alpha$ in the primitive Gaussian $p$, and $B$ is a normalization constant.
The \gls{cgto} for atom $I$ of species $\alpha_I$ at position $R_I$ is therefore
\begin{equation}
    \phi_{nlm}^{I}(\mathbf{r}) = R_{nl}^{\alpha_I}(|\mathbf{r} - \mathbf{R}_I|) Y_l^m(\theta,\phi).
\end{equation}
For periodic systems, we can write approximate Bloch states as Bloch sums of \glspl{cgto}
\begin{equation}
    \phi_{\nu\mathbf{k}}(\mathbf{r}) = \sum_{\mathbf{R}} e^{i\mathbf{k}\cdot\mathbf{R}}\phi_{nlm}^{I}(\mathbf{r}-\mathbf{R})
\end{equation}
with band index $\nu$ capturing the \gls{cgto} indices $n$, $l$, $m$, and $I$, and crystal wave vector $\mathbf{k}$.
We define the electronic-structure map as a simple decoration of the atomic positions with the \glspl{cgto} of the corresponding species, which requires no significant computational effort.

A simple yet descriptive one-electron integral is the kinetic energy, which can be applied to the Bloch sums
\begin{equation}
    T_{\nu\mu\mathbf{k}} = \matrixel{\phi_{\nu\mathbf{k}}}{-\frac{\hbar^2}{2}\nabla^2}{\phi_{\mu\mathbf{k}}}
\end{equation}
and subsequently diagonalized.
The density of kinetic energy eigenvalues per unit volume can then be calculated using Gaussian smearing.
Volume normalization ensures that the spectra for unit cells and supercells are identical, which is the desired behavior in solids when predicting intrinsic properties such as structural similarity or electronic band gap.
The kinetic energy \gls{sorep} features are therefore
\begin{equation}
    \label{eq:gaussian_dos}
    x_{i} = \frac{1}{V} \frac{1}{N_{\nu} N_{\mathbf{k}}} \sum_{\nu \mathbf{k}}\exp \left[ \frac{-[(E_i - \lambda_{\nu \mathbf{k}}) / (k_B T_s)]^2}{\sqrt{\pi}} \right]
\end{equation}
where $T_s$ is a smearing temperature, $E_i$ are uniformly-spaced energies running from $E_{\mathrm{min}}$ to $E_{\mathrm{max}}$, and $\lambda_{\nu \mathbf{k}}$ are the kinetic energy eigenvalues.

\begin{figure}
    \centering
    \includegraphics[width=1.0\linewidth]{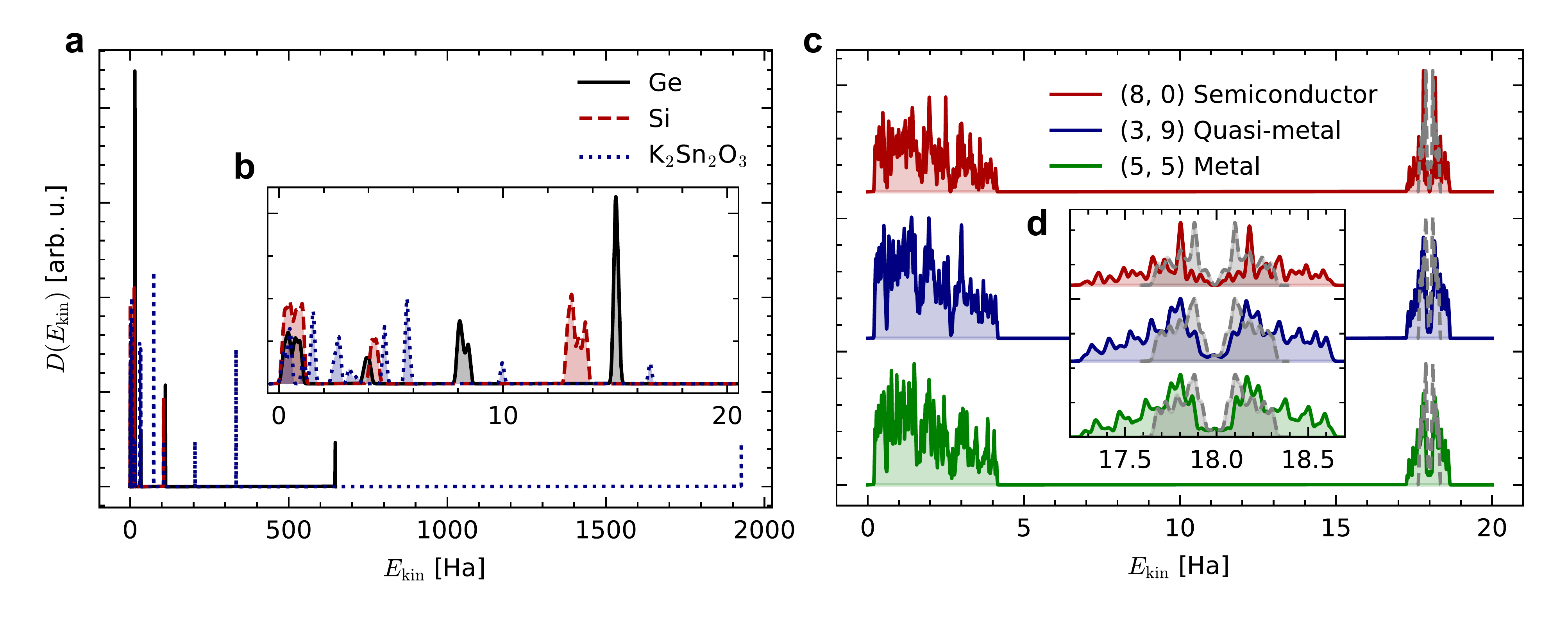}
    \caption{\textbf{a} Kinetic energy \glspl{sorep} of diamond-structure Ge and Si along with the transparent conducting oxide K$_2$Sn$_2$O$_3$.
    The width of the spectral range ($\approx 2000$ Ha) is dominated by tightly-bound semi-core orbitals of K$_2$Sn$_2$O$_3$.
    \textbf{b} The two elemental solids display qualitatively similar features at low energies ($\lesssim 20$ Ha).
    \textbf{c} Kinetic energy \glspl{sorep} for three carbon nanotubes of different chiralities, overlaid with $p_z$ tight-binding densities of state (dashed lines, shifted by +18 Ha to be similarly centered).
    \textbf{d} The transition from metallic to insulating configurations is correlated with a disappearance of the kinetic \gls{dos} in the (3,9) and (0,8) configurations around approximately 18 Ha.}
    \label{fig:kin_dos}
\end{figure}

\cfref{fig:kin_dos}a shows the kinetic energy \glspl{sorep} for silicon, germanium, and K$_2$Sn$_2$O$_3$ calculated using a customized version of the \gls{ano} \gls{cgto} basis set~\cite{Pritchard_2019,Feller_1996,Schuchardt_2007,Roos_2004a,Roos_2004b,Roos_2005a,Roos_2005b,Roos_2008,Widmark_1990} domain which covers the full energy spectrum.
This custom basis set is obtained from the relativistic \gls{ano}-type orbitals ~\cite{Roos_2004a,Roos_2004b,Roos_2005a,Roos_2005b,Roos_2008,Widmark_1990} known as \gls{ano}-RCC, as available on the Basis Set Exchange library~\cite{Pritchard_2019,Feller_1996,Schuchardt_2007}, where we keep only the orbitals corresponding to the smallest closed-shell configuration.
As one might expect, the \glspl{sorep} for the two elemental solids look qualitatively quite similar, mirroring their similar structural electronic properties.
When comparing these materials to a more complex system with heavier elements, like K$_2$Sn$_2$O$_3$ as shown, we observe a rapidly growing spectral range due to the increasingly highly-localized nature of the additional tightly-bound (semi-)core orbitals.
In order to improve the features for this purpose, one might consider modifying the operator by adding a nuclear potential or changing the final representation to one that compresses the spectrum more than the \gls{dos}.
For example, the \gls{dos} could be truncated at the energy where the cumulative \gls{dos} is some fraction of its maximum, following the observation that only a small amount of the \gls{dos} is at high energy.
Applying such a cutoff would yield features similar to those pictured in \cfref{fig:kin_dos}b.
The underlying complication is not only that the spectral range is large but that it differs between chemistries in a way that makes it difficult to select the most important region(s) of the \gls{dos} beyond the intuition that low-curvature, relatively delocalized, and thus low-kinetic-energy states are more chemically meaningful than high-curvature, highly-localized, high-kinetic-energy ones.
However, for systems of the same or similar compositions, such as the \glspl{cnt} shown in \cfref{fig:kin_dos}c, this filtering is more intuitive.
Here, we compare the kinetic \glspl{sorep} as above for \glspl{cnt} of varying electronic character and find that the features around 18 Hartree, as seen in \cfref{fig:kin_dos}d, are quite similar to the $p_z$ tight-binding \glspl{dos} \cite{Yang_1999,Seol_2006} (shown in gray) and exhibit a gap forming for the semiconducting (0, 8) configuration.
We conclude that the kinetic features are well suited to comparing such compositionally similar materials, like in molecular dynamics, metallic alloys, or elemental systems with many allotropes.

\begin{figure*}
    \centering
    \includegraphics[width=0.95\linewidth]{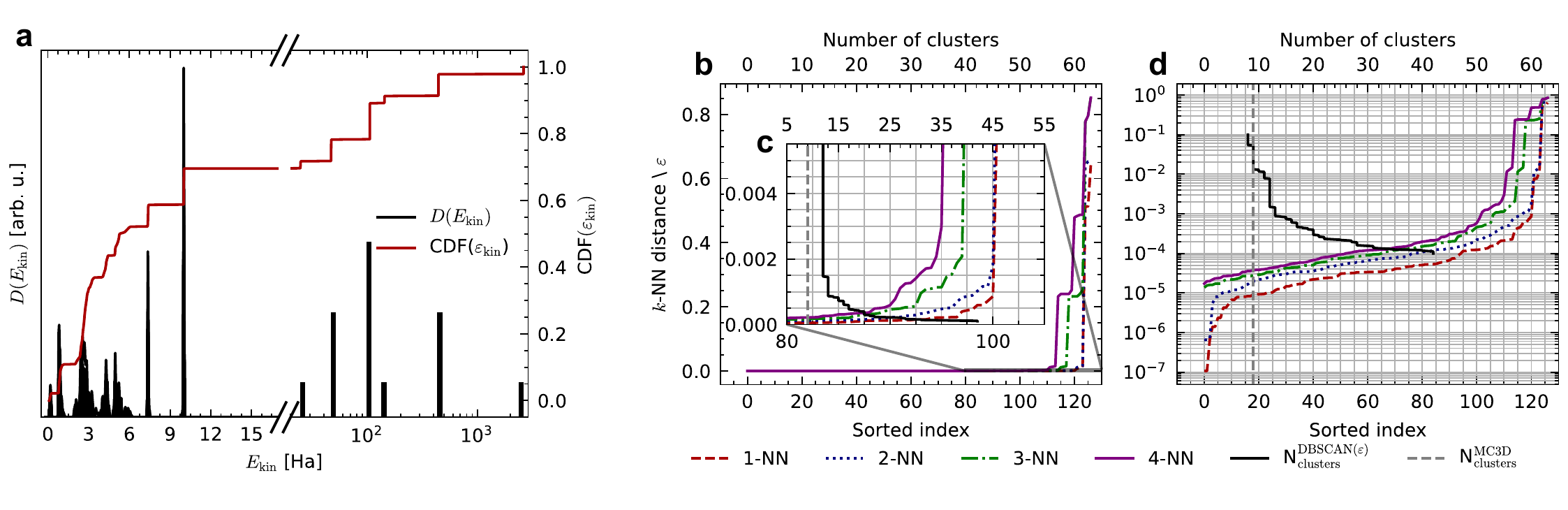}
    \caption{
        \textbf{a} Density and cumulative distribution function of kinetic eigenvalues for all BaTiO$_3$ structures considered. Below 15 Hartree, the \gls{dos} is computed with Gaussian smearing; above, a logarithmically binned histogram is shown.
        \textbf{b} Sorted $k$-nearest neighbor distance curves on a linear scale and \textbf{d} on a semi-log scale display \textbf{c}, \textbf{d} an ``elbow'' at optimal values of the \gls{dbscan} neighborhood radius parameter $\varepsilon$.
        \gls{dbscan} models are fit for $\varepsilon$ values within the elbow region (approximately $1 \times 10^{-4}$ to $1 \times 10^{-1}$), and the corresponding number of clusters is shown in solid black.
        The number of structure groups determined by the \gls{mc3d} procedure is overlaid in dashed gray for reference.
        An optimal choice of $\varepsilon$ exists in the region of relatively stable clustering around $1 \times 10^{-3}$ to $1 \times 10^{-1}$.
    }
    \label{fig:batio3_combined}
\end{figure*}

One application in this regard is identifying unique structures of a fixed stoichiometry in a large database.
For illustration, we have selected the relaxed geometries for 127 BaTiO$_3$ entries in the \gls{mc3d} \cite{Huber_2022} feedstock and compare against the uniqueness analysis conducted by the \gls{mc3d} developers using the \texttt{pymatgen} structure matcher \cite{PingOng_2013}.
To find an appropriate energy range for featurization, we first analyze the inexpensive-to-compute cumulative distribution of kinetic eigenvalues across all structures, shown in \cfref{fig:batio3_combined}a.
This cumulative distribution function increases in quasi-discrete steps at energies higher than approximately 7 Hartree.
Past this point, the density of kinetic states is likely dominated by tightly-bound (semi-)core states, which are likely uninformative.
The \gls{dos} computed up to 15 Hartree, also in \cfref{fig:batio3_combined}a, is indeed sparse, highly peaked, and therefore likely related to highly-localized Ba and Ti \glspl{cgto} above about 6.5 Hartree.
The final \gls{sorep} features are therefore computed from 0 to 6.5 Hartree using Gaussian-type smearing with a width of 0.03 Hartree sampled at 1024 equally-spaced energies.

\begin{figure}
    \centering
    \includegraphics[width=0.95\linewidth,trim={0 5cm 1cm 5cm},clip]{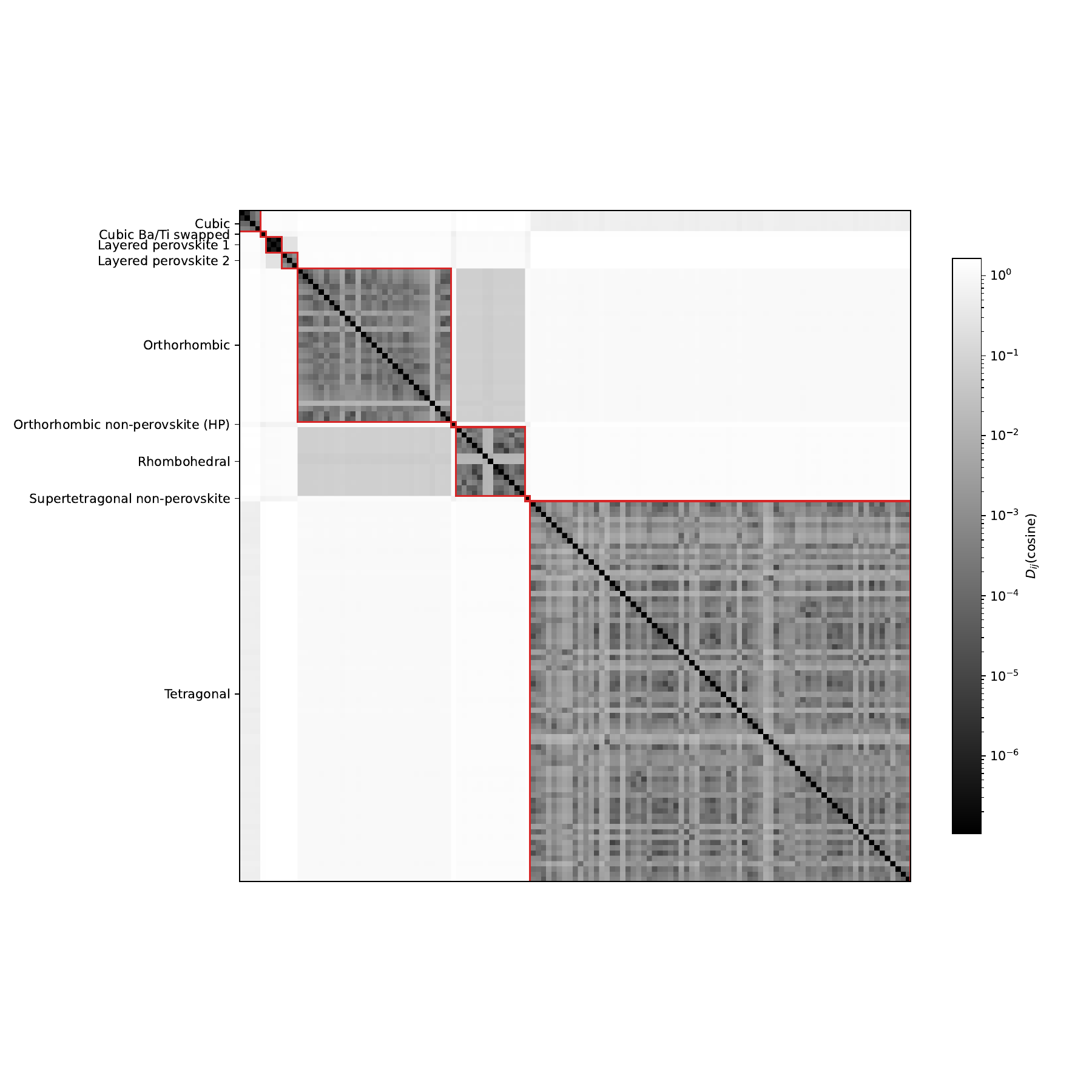}
    \caption{Cosine distance matrix relating kinetic \glspl{sorep} for 127 BaTiO$_3$ structures with columns sorted by \gls{mc3d} classification. Some outliers stand out visually in the orthorhombic and rhombohedral groups alongside possible outliers in the tetragonal block.}
    \label{fig:batio3_kin_cos_distances}
\end{figure}

To determine a set of unique prototype structures, the \gls{sorep} features are clustered using the \gls{dbscan} \cite{Ester_1996} algorithm, which has two parameters: the minimum number of samples required to create a cluster $N_{min}$, and a neighborhood radius  $\varepsilon$.
\gls{dbscan} performs its clustering based on the distances between data points, not the data themselves; here, we use the cosine distance $d_{cos}(\mathbf{x}_i, \mathbf{x}_j) = \frac{\mathbf{x}_i \cdot \mathbf{x}_j}{||\mathbf{x}_i||||\mathbf{x}_j||}$.
The distance matrix, sorted by the \gls{mc3d} grouping, is shown in \cfref{fig:batio3_kin_cos_distances}.
Here, there are three groups containing only one structure each (an orthorhombic non-perovskite, a super-tetragonal non-perovskite, and an erroneous structure with the positions of barium and titanium swapped) along with six groups with multiple members (two layered perovskites, and the four standard polymorphs).
However, it can be seen that some groups (notably those labeled as orthorhombic and rhombohedral) contain structures which are relatively unlike the rest of their respective clusters.

As we expect to find some completely unique structures with no duplicates in the database, we set $N_{min} = 1$ for the following investigation.
To determine $\varepsilon$, we inspect the sorted \gls{knn} distances (\cfref{fig:batio3_combined}b,c,d) and find an ``elbow'' in the curve, which indicates a domain of reasonable values, from approximately $1\times10^{-4}$ to $1\times10^{-1}$.
Correlating these values of $\varepsilon$ with the number of clusters found by \gls{dbscan}, it can be seen that within this region of the \gls{knn} distance curve, the number of unique structure groups ranges from more than 50 to 8 with increasing $\varepsilon$, compared to the 9 groups predicted by the \gls{mc3d} procedure.
However, the clustering is highly sensitive to small changes in the neighborhood radius for $\varepsilon < 9\times10^{-4}$, so we focus on values in the range $[9\times10^{-4},1\times10^{-1}]$ which yield between 12 and 8 groups.
The 9 clusters produced for $\varepsilon$ in the range $[0.014, 0.05]$ are identical to those from the \gls{mc3d} as determined using \texttt{pymatgen}'s structure matcher.
To understand how the clustering at lower and higher values of $\epsilon$ differs within the focus region, we describe the process of cluster merging as $\varepsilon$ increases in \ctref{tab:batio3_kin_dbscan_clusters}.
Of the 13 groups generated at low $\varepsilon$, most are in agreement with the reference groups aside from a splitting of the tetragonal, orthorhombic, and rhombohedral clusters, where the visual outliers seen in \cfref{fig:batio3_kin_cos_distances} are separated.

\begin{table*}
    \centering
    \begin{tabular}{l|cccccc}
 & \multicolumn{6}{c}{$\varepsilon$}\\
        \textbf{Group}                     &~$1\times10^{-3}$&~$2\times10^{-3}$&~$1\times10^{-2}$&~$1.2\times10^{-2}$&~$2\times10^{-2}$&~$7\times10^{-2}$\\
        \hline \hline
        Tetragonal                         & 69               & \textbf{72}      &  -               &  -                 & -                &72                \\
        ~~~$^1$                            &  3               & $\uparrow$       &                  &                    &                  &                  \\
        Orthorhombic                       & 28               &  28              &  28              &  \textbf{29}       & -                &$\downarrow$      \\
        ~~~$^2$                            &  1               &   1              &  1               &  $\uparrow$        &                  &                  \\
        Rhombohedral                       & 11               &  11              &  11              &  11                & \textbf{13}      &$\downarrow$      \\
        ~~~$^3$                            &  1               &   1              &  \textbf{2}      &  2                 & $\uparrow$       &                  \\
        ~~~$^4$                            &  1               &   1              &  $\uparrow$      &                    &                  &                  \\
        Orthorhombic or Rhombohedral       &                  &                  &                  &                    &                  &\textbf{42}       \\
        \hline
        Cubic                              &  4               &  -               &  -               &  -                 & -                &4                 \\
        Cubic Ba/Ti swapped                &  1               &  -               &  -               &  -                 & -                &1                 \\
        Layered perovskite 1               &  3               &  -               &  -               &  -                 & -                &3                 \\
        Layered perovskite 2               &  3               &  -               &  -               &  -                 & -                &3                 \\
        Orthorhombic non-perovskite        &  1               &  -               &  -               &  -                 & -                &1                 \\
        Supertetragonal non-perovskite     &  1               &  -               &  -               &  -                 & -                &1                 \\
    \end{tabular}
    \caption{
        Progression of \gls{dbscan} clusters with varying neighborhood radius parameters $\varepsilon$.
        As the neighborhood radius is increased, clusters merge, agreeing with the \gls{mc3d}'s structure-based grouping for values from $2\times10^{-2}$ to approximately $5\times10^{-2}$, after which meaningfully distinct clusters combine.
}
    \label{tab:batio3_kin_dbscan_clusters}
\end{table*}

We conclude from this analysis that the kinetic \gls{sorep} features have the ability to capture and describe subtle structural differences in polymorphs of complex materials, similar to structure-based methods like the \texttt{pymatgen} structure matcher.
However, significant weight is given specifically to structural changes that strongly affect the electronic structure, as seen in the \glspl{cnt}.
This simple and efficient example serves as a good case study for how one might approach crafting and electronic-structure featurization under the \gls{sorep} framework with a quite restrictive efficiency constraint.
As mentioned above, more complex learning problems do often require more expressive descriptors, so next we consider constructing a featurization for such a situation.

\subsection*{Single-shot DFT SOREP}
\label{sec:ssdft}
For applications such as screening large and diverse databases of materials, a representation is required that is rich enough to describe and compare any chemical composition but computationally efficient enough to be applied to tens of thousands of systems (containing up to \~{}100 atoms each).
To guide the development of a \gls{sorep} fit for this application, we consider as a use case a \gls{ml}-accelerated screening for \glspl{tcm} in the \gls{mc3d}.
\Glspl{tcm} are characterized by band gaps wide enough to allow for transparency across the visible spectrum, high mobility of charge carriers, and the ability to inject these carriers via n- or p-type doping.
Most screening studies for these materials focus initially on approximating the first two properties via high-throughput \gls{dft} band-structure calculations \cite{Hautier_2013,Hautier_2014,WoodsRobinson_2018}.
From these calculations, the \gls{dft}-PBE \cite{Perdew_1996} band gap and approximate electron and hole effective masses are used as figures of merit.
Our aim is to define a featurization method descriptive enough to reproduce a classification based on effective masses and band gaps at a fraction of the cost.
More concretely, we target an order of magnitude speedup compared to self-consistent \gls{dft} calculations; otherwise, it would be more efficient and practical to perform a traditional screening.
Using these guiding principles, we propose a \gls{sorep} method based on a single-shot (i.e. non-self-consistent) \gls{dft} calculation of a superposition of pseudo-atomic valence charge densities $\tilde{\rho}$ and a linear combination of \glspl{pao} $\chi_{nl}$ taken from pseudopotentials from the \gls{sssp} library \cite{Lejaeghere_2016,Prandini_2018}.
These pseudo-atomic quantities are exact matches to the all-electron quantities of an isolated atom outside a small pseudization radius and as such represent a reasonable guess of the true ground-state wavefunction and charge density of the chemically-active electrons of each element.
Using a locally modified copy of the Quantum ESPRESSO \cite{Giannozzi_2009} \texttt{pw.x} code, the pseudo-atomic orbitals (provided on a real-space radial grid) are transformed into Bloch orbitals.
The Kohn-Sham \gls{dft} Hamiltonian and orbital overlap matrices are then calculated non-self-consistently from the potential derived from the superposition of atomic densities
\begin{align}
    H_{\nu\nu\prime\mathbf{k}} &= \matrixel{\chi_{\nu\mathbf{k}}}{\hat{H}}{\chi_{\nu\prime\mathbf{k}}} \\
    S_{\nu\nu\prime\mathbf{k}} &= \braket{\chi_{\nu\mathbf{k}}}{\chi_{\nu\prime\mathbf{k}}}
\end{align}
where
\begin{equation}
    \hat{H} = \hat{T} + \hat{V}_{H[\tilde{\rho}]} + \hat{V}_{xc[\tilde{\rho}]} + \hat{V}_{ext} + \hat{V}_{PS}.
\end{equation}
The Hamiltonian matrix then is diagonalized exactly (i.e. non-iteratively) on the basis of the pseudo-atomic orbitals to find the eigenvalues $\varepsilon_{n\mathbf{k}}$ and eigenstates $\psi_{n\mathbf{k}}$ at each $\mathbf{k}$-point
\begin{equation}
    H_{\mathbf{k}} \psi_{n\mathbf{k}} = \varepsilon_{n\mathbf{k}} S_{\mathbf{k}} \psi_{n\mathbf{k}},
\end{equation}
yielding a $\mathbf{k}$-resolved eigenspectrum, i.e.~a band structure.
With respect to the kinetic energy operator used above, the Kohn-Sham Hamiltonian is well-behaved due to the inclusion of potential terms, with a meaningful Fermi energy and band extrema that can be leveraged as anchoring points.
Finally, the features are the \gls{dos} calculated with Gaussian smearing as in \ceref{eq:gaussian_dos} where the discretization over energies $E_i$, and smearing temperature $T_s$ are taken as parameters of the featurization.

\begin{figure}
    \centering
    \includegraphics[width=0.95\linewidth]{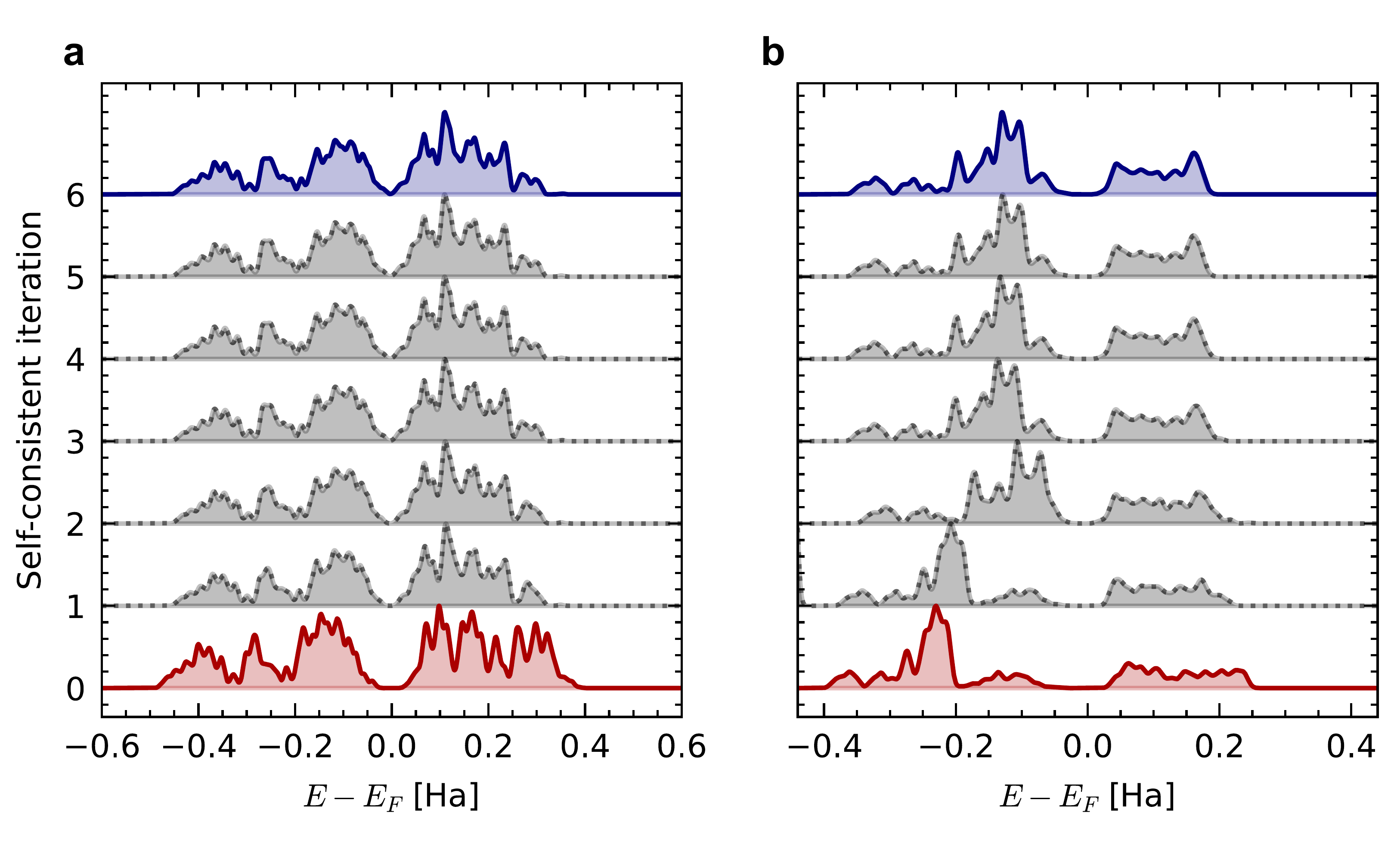}
    \caption{
        Densities of states for \textbf{a} silicon and \textbf{b} K$_2$Sn$_2$O$_3$ across self-consistent iterations.
        ``0'' iterations corresponds to a single-shot \gls{sorep}, while in both cases, the uppermost \gls{dos} corresponds to the self-consistent DFT ground state.
        The \gls{sorep} for silicon is remarkably similar to the converged \gls{dos}, while the features for the more complex K$_2$Sn$_2$O$_3$ are further from the ground-state solution.
    }
    \label{fig:qe_qnscf_dos}
\end{figure}

\cfref{fig:qe_qnscf_dos} shows the evolution of the \gls{dft} \gls{dos} with respect to electronic self-consistent step for a reference semiconductor, elemental silicon, and the transparent conducting oxide K$_2$Sn$_2$O$_3$.
Remarkably, the qualitative shape of the \gls{dos} is quite similar to the fully converged calculation, especially around the Fermi level.
Importantly for the application of screening \glspl{tcm}, the shape of the \gls{dos} at the conduction and valence band edges is well-reproduced in the single-shot \glspl{sorep}, so the features contain, to some extent, reliable information related to the electron and hole effective masses.

\subsection*{Accelerated TCM screening}
\label{sec:screening}

\begin{figure}
    \centering
    \includegraphics[width=0.95\linewidth]{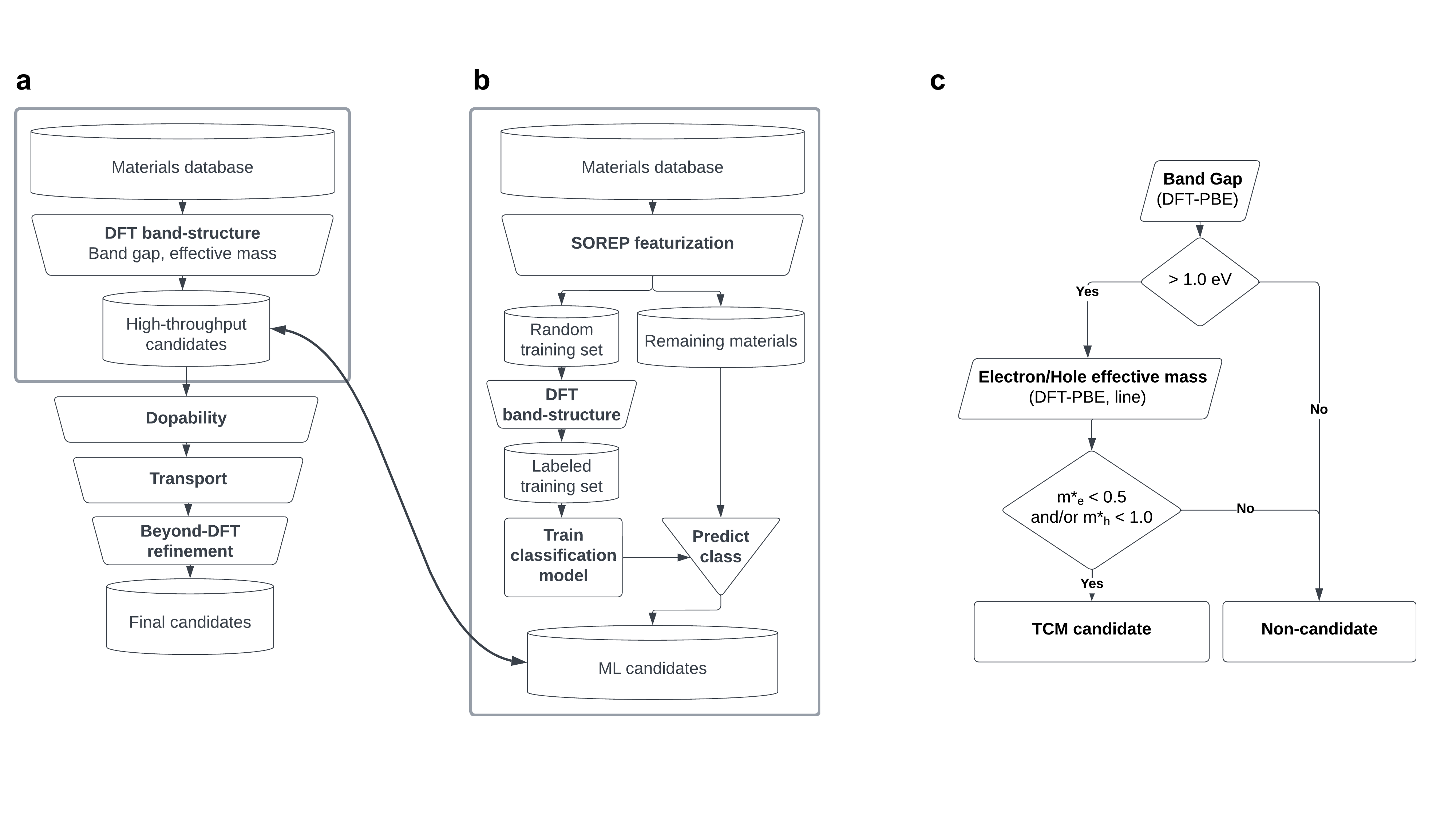}
    \caption{
        \textbf{a} Flow diagram of a high-throughput screening for \glspl{tcm}, where the boxed section corresponds to the steps accelerated by \gls{sorep}-based \gls{ml}.
        \textbf{b} \gls{ml}-accelerated procedure: random materials from the database are used to train a classification model which labels the remaining materials. With a perfect classification model, the ``\gls{ml} candidates'' in \textbf{a} would correspond exactly to the ``high-throughput candidates'' in \textbf{b}.
        \textbf{c} \gls{tcm} screening criteria based on band gaps and hole and electron effective masses.
    }
    \label{fig:screening}
\end{figure}

A generalized procedure for screening transparent conducting materials is shown in \cfref{fig:screening}a, and consists of two broad steps: high-throughput \gls{dft} calculations (boxed) and low-throughput refined calculations.
The most computationally expensive step is running \gls{dft} band-structure workflows for every material in the database, numbering often in the tens of thousands.
These calculations are used to find the band gap and effective mass (proxies for transparency and conductivity respectively) following the criteria shown in \cfref{fig:screening}c.
To accelerate the procedure, we featurize the database using the single-shot \gls{dft} \glspl{sorep} and construct a classification model which predicts which materials are likely to meet the \gls{dft}-based screening criteria (\cfref{fig:screening}b).
This approach significantly reduces computational cost by performing \gls{dft} band-structure calculations only on a subset of the entire database, including, if available, some known \glspl{tcm} in order to construct an \gls{ml} model used to screen the rest of the database.
Note that this approach is more general than the case of \glspl{tcm}; screening studies for many materials classes and properties could follow a similar procedure.

As this is a validation study and not a true screening, we take the \gls{mc3d}, a curated database of relaxed three-dimensional crystal structures, for ground-truth \gls{dft} simulation results.
Using the high-symmetry line band structures from the \gls{mc3d}, all materials are classified as \gls{tcm} candidates or non-candidates based on the criteria shown in \cfref{fig:screening}c. The filter, adapted from general guidelines outlined by Woods-Robinson \etal~\cite{WoodsRobinson_2018}, select candidate materials first by having a \gls{gga} electronic band-gap wider than 1 electron-volt. Then, if the material meets either an electron or hole effective mass condition (or both), it is considered a candidate material; otherwise, it is labeled as a non-candidate.

The band gap is simply calculated as the difference between the \gls{cbm} and \gls{vbm} $E_g = E_{\mathrm{CBM}} - E_{\mathrm{VBM}}$ found in the band dispersion from a self-consistent \gls{dft} calculation.
The electron effective masses are approximated from band structures computed along high-symmetry lines provided by SeeKPath \cite{Hinuma_2017} using the so-called ``line effective mass'' of Hautier \etal~\cite{Hautier_2014}:
\begin{equation}
    \label{line_eff_mass}
    \frac{1}{m^*_{e,\mathrm{line}}} = \max_{\alpha} \left[\frac{
        \sum_{n \in \mathrm{CB}} \int_{k_{\alpha a}}^{k_{\alpha b}} -\frac{\partial^2}{\partial k_\alpha^2} \varepsilon_{n}(k_\alpha) \theta_e(\varepsilon_n(k_\alpha)) dk_\alpha
    }{
        \sum_{n \in \mathrm{CB}} \int_{k_{\alpha a}}^{k_{\alpha b}} \theta_e(\varepsilon_n(k_\alpha)) dk_\alpha
    }\right]
\end{equation}
where the maximum is taken over high-symmetry lines $\alpha$, the sum is over conduction bands, and $\theta$ is the Fermi-Dirac distribution at 300K:
\begin{equation}
    \theta_e(E) = \left[  \exp \left( \frac{E - E_{\mathrm{CBM}}}{k_{B} T_{300K}} \right) + 1 \right]^{-1}.
\end{equation}
Hole effective masses are approximated similarly by exchanging $E - E_{\mathrm{CBM}}$ in the Fermi-Dirac distribution for $E_{\mathrm{VBM}} - E$ and summing over valence, rather than conduction, bands in \ceref{line_eff_mass}.

These labeled data are then used to train a \gls{rfc} to predict the \gls{dft}-derived binary classification based on the \gls{sorep} features; the \gls{rfc} model is chosen for its simplicity and interpretability.
Unlike neural network models, which are often quite opaque, the binary decisions in each of the trees of the random forest are easily understood as conditions on the \gls{dos} at particular energies, and the model can provide a measure of the importance of each of the input features.
Because the input features have clear physical meanings, the results of the model training not only provide classification predictions but also useful feedback for improving the \gls{sorep} features if necessary.

\subsection*{Featurization}
\label{sec:featurization}

Single-shot calculations of 30,054 of 35,240 materials available in the \gls{mc3d}, ranging in size from 1 to 80 atoms, are completed within a 45 core-minute per material limit, 22,200 of which had available band-structure data at the time of retrieval.
2,527 systems of the 22,200 with band structures are classified as possible \glspl{tcm}, while the remaining 19,673 do not meet the band-gap or effective mass criteria.
This yields a relatively strongly imbalanced data set with 7.78 negative samples per positive sample.

We investigate three different \gls{sorep} parameterizations, focusing on different aspects of the \gls{dos} in a similar spirit to a weighted \gls{dos} fingerprint \cite{Kuban_2022}.
The first is centered around the valence band maximum and samples a small amount of the valence bands and 6 eV into the band gap / conduction bands.
These features are designed with selecting hole-conducing \glspl{tcm} in mind; by fixing the \gls{dos} at the \gls{vbm} to a specific feature, the model has a higher likelihood to learn to distinguish materials with high density of states, and likely high effective mass, at that point.
Including 6 eV above the \gls{vbm} should also provide enough information about the band gap so that insulators may be distinguished from conductors.
For most insulators, this energy range should also provide a glimpse at the bottom of the conduction bands.
The second set of parameters yields a standard Fermi-level-centered \gls{dos} with, sampled on a range of $\pm$5 eV, allowing the valence and conduction bands to be captured in any material with a band gap less than 10 eV.
For insulating materials, the Fermi level is taken as the mid-gap energy so that the valence and conduction are equally well represented.
To determine whether and how much the model relies on information about the band gap, a third parameterization removes energies within the band gap by ``scissoring'' them away.
All eigenvalues above the conduction band maximum are shifted down by the width of the gap minus a small tolerance factor of three smearing widths.
An energy range of 2 eV is sampled into the valence and conduction bands, providing a similar sampling of the bands as the Fermi level-centered features for a 6 eV gap material, albeit at a higher resolution due to the constant number of energies taken on the range.

\begin{figure}[ht!]
    \centering
    \includegraphics[width=0.95\linewidth]{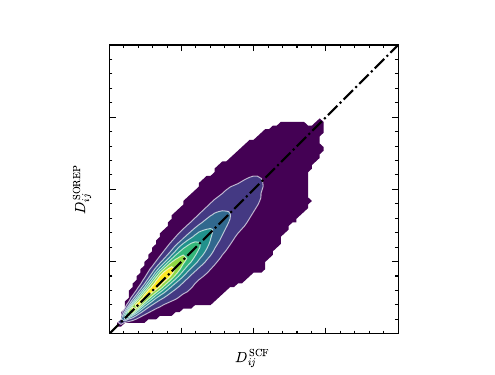}
    \caption{Parity plot comparing Euclidean distances among SCF \glspl{dos} and corresponding distances among single-shot \gls{dft} \glspl{sorep}. The distances are highly correlated between the two sets of features, suggesting that the space of single-shot densities of states is not too different from that of fully self-consistent \glspl{dos}, which have been previously employed for materials cartography.}
    \label{fig:distance_parity_fermi}
\end{figure}

As shown in Refs.~\citeonline{Isayev_2015} and \citeonline{Geilhufe_2018}, distance metrics based on the self-consistent \gls{dft} \gls{dos} can be used in practice to both map out the space of electronic structures and to search for materials with complex electronic properties.
To more rigorously investigate the information content of the \glspl{sorep} described above, we observe the correlation between the properties of the Fermi-level centered single-shot \gls{dft} \gls{sorep} features and identically sampled SCF densities of state from the \gls{mc3d}.
\cfref{fig:distance_parity_fermi} shows a strong correlation between self-consistent \gls{dft} and \gls{sorep} Euclidean distances across the database, confirming that this featurization not only contains physically relevant information but that it yields a topologically similar space to true self-consistent densities of state.
This confirmation opens the door to applying \glspl{sorep} to materials cartography and other further investigations into electronic-structure space and its dimensionality using tools like DADApy \cite{Glielmo_2022}.

\subsection*{Model training and performance}
\label{sec:performance}
Using the three \gls{sorep} parameterizations described above, balanced \gls{rfc} models which correct for heavily imbalanced training data, like we have here, are trained using a gridded parameter cross-validation with $k=5$ folds on a range of train-test splits.
On a 14-core workstation, training all the models for all train-test splits and all \gls{sorep} features can be done in less than an hour.

\begin{figure}
    \centering
    \includegraphics[width=0.95\linewidth]{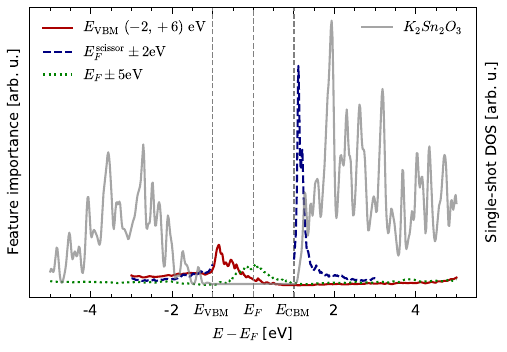}
    \caption{
        Random forest classifier feature importances for models trained on 65\% of the data ($N=14430$), smoothed using a Gaussian filter ($\sigma=1.2dE$) for clarity.
        In gray, the Fermi-centered \gls{sorep} of $K_2Sn_2O_3$ is shown for context as an example set of features.
        For the \gls{vbm}-centered \glspl{sorep}, the model finds features just above the \gls{vbm} to be most important; for Fermi-level-centered features, $\pm\approx0.6$~eV around the Fermi level are most important, with a bias towards the \gls{cbm}; and for ``scissor'' features, the conduction band edge is by far the most important.
    }
    \label{fig:feature_importance}
\end{figure}

A first question to ask after training the models is: what features do the models find to be most useful in their predictions?
\cfref{fig:feature_importance} shows the feature importances for the three types of \gls{sorep} in models trained on 65\% of the database (the maximum training fraction investigated).
The \gls{vbm}-centered model gives significant importance to the valence bands and exhibits a strong peak just above the \gls{vbm} and into the gap (if present), likely to probe the shape of the decay of the \gls{dos} at the band edge and the width of the gap.
This follows the physical intuition used in constructing these features: the shape of the bands around the \gls{vbm} and information about the band gap are both important for classifying materials as \glspl{tcm} given the criteria imposed.
By contrast, the features with the gap ``scissored'' out show a much stronger peak around the conduction band edge, with relatively low importance given to the valence band edge, both of which have fixed positions within the feature vector (either side of the center of the features).
The full Fermi-centered features show a combination of the behaviors above: a peak at the Fermi level and the surrounding energies give information about the presence of a gap and possibly about the shape of the DOS at band edges if close enough to the Fermi level.

\begin{table*}
    \centering
    \begin{tabular}{c|c|c||c|c|c|c}
        $E_0$             &       $E_{\mathrm{min}}$ (eV)&       $E_{\mathrm{max}}$ (eV)&    True Positive&   False Positive&    True Negative&   False Negative\\
                          &                              &                              &           (Rate)&           (Rate)&           (Rate)&           (Rate)\\
        \hline \hline
        $E_{\mathrm{VBM}}$&                            -2&                            +6&            1,903&              600&           14,546&            4,929\\
                          &                              &                              &          (0.760)&          (0.240)&          (0.747)&          (0.253)\\
        \hline
          $E_{\mathrm{F}}$& $E_{\mathrm{VBM}} + 3T_s - 2$& $E_{\mathrm{CBM}} - 3T_s + 2$&            1,897&              606&  \textbf{16,015}&  \textbf{3,460}\\
                          &                              &                              &          (0.758)&          (0.242)& \textbf{(0.822)}& \textbf{(0.178)}\\
        \hline
          $E_{\mathrm{F}}$&                            -5&                            +5&   \textbf{2,128}&     \textbf{375}&           14,711&            4,764\\
                          &                              &                              & \textbf{(0.850)}& \textbf{(0.150)}&          (0.755)&          (0.245)\\
    \end{tabular}
    \caption{True and false positive counts and rates for models trained on 1\% of the data (N = 222). The best performing features are shown in bold; standard Fermi level-centered features achieve the best true positive rate, while Fermi level-centered features disregarding energies within the gap discern true negatives best. Because of their dominating population, the performance of the features is most heavily influenced by their ability to correctly label negative (non-\gls{tcm}) samples.}
    \label{tab:confusion}
\end{table*}

Although the feature importances tell quite different stories for each of the different \glspl{sorep}, confusion matrices shown in \ctref{tab:confusion} confirm that the relative performance of the features is comparable.
In broad terms, the models trained using 1\% of the database each achieve true positive rates of above 75\% and up to 85\% with corresponding false negative rates of between 18\% and 25\%.
These quantities are important measures of how well a given classification model can accelerate a screening: a high false positive rate would mean that significant calculation time is wasted on false leads, while a low true positive rate would signal that many good candidates go overlooked.
\begin{figure}[ht!]
    \centering
    \includegraphics[width=0.95\linewidth]{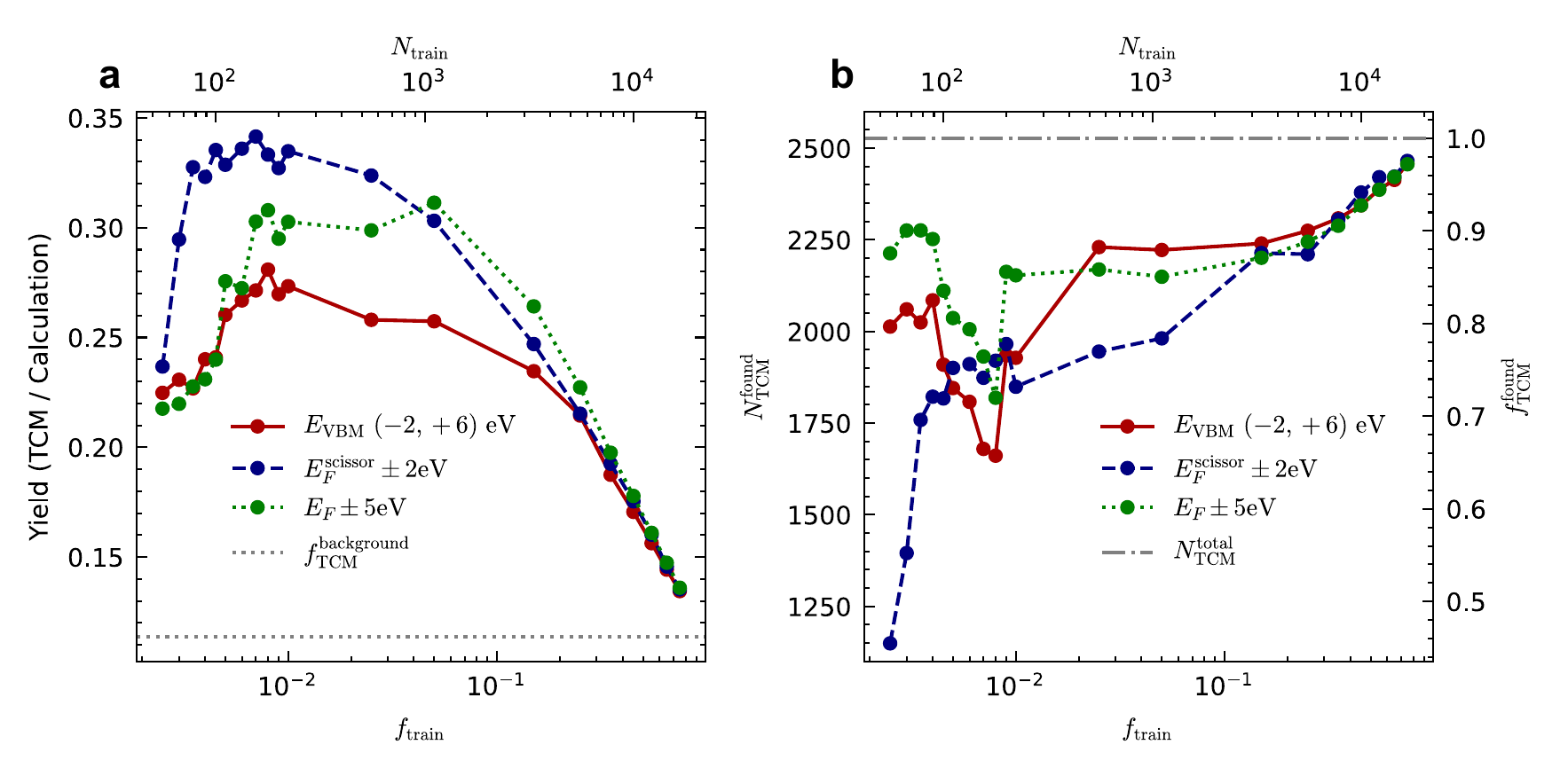}
    \caption{
        \textbf{a} Yield curves for the three feature methods, i.e. the fraction of calculations performed for both training and evaluation which yield \gls{tcm} candidates.
        \textbf{b}: Number of \gls{tcm} candidates found during both training and evaluation w.r.t. the total number present in the database.
        The yield increases rapidly at low training fractions due to a reduction in false positive rates, while the corresponding number of \glspl{tcm} found decreases due to worsening true positive rates.
        Above approximately 1\% training fraction (N$\approx$200), model performance plateaus while the role of training data production becomes more significant.
    }
    \label{fig:yield}
\end{figure}
The effects of these factors are most clearly shown in \cfref{fig:yield}, which shows the yield of promising \gls{tcm} candidates found per calculation during creation of the training data and in confirming the predictions of the model:
\begin{equation}
    \label{eq:yield}
    \mathrm{yield} = \frac{
        N_{\mathrm{TP}} + N_{\mathrm{train}} * f_{\mathrm{TCM}}^{\mathrm{background}}
        }{
        N_{\mathrm{FP}} + N_{\mathrm{train}} + N_{\mathrm{TP}}
        }.
\end{equation}
$N_{\mathrm{TP}}$ is the number of true positives (\glspl{tcm} labeled as such), $N_{\mathrm{FP}}$ the number of false positives (non-\glspl{tcm} labeled as \glspl{tcm}), $N_{\mathrm{train}}$ the number of training samples, and $f_{\mathrm{TCM}}^{\mathrm{background}}$ the fraction of materials in the database which meet the \gls{tcm} criteria.
Counterintuitively, in the case of screening where producing training data finds candidates at the background rate, decreasing the training fraction actually increases the yield rate.
This is true as long as the model has a better precision than the fraction of positive samples in the database.
The median precision of all the models trained is 0.33, three times better than the background occurrence of \glspl{tcm} in the database at 0.11.
The fractional yield can still hide significant decreases in total yield, as seen in \cfref{fig:yield}b, where the total yield for models trained on fewer than 200 materials is much lower than the fractional yield might suggest.
This difference is due to a breakdown in the representativity of the training data, which contain too few (if any) positively labeled samples, leading the model to overly favor negative predictions.

The most efficient model according to \cfref{fig:yield}a uses the ``scissored'' features and is trained on 0.5\%-1.5\% of the database, or between 100 and 300 materials.
After training on 222 materials, the \gls{rfc} predicts 1,897 true positives and 3,460 false positives, requiring in total 5,579 SCF and band-structure calculations, or 25\% of the database.
In turn, it finds 76\% of the \gls{tcm} candidates present in the database.
In terms of total number of \glspl{tcm} found, the valence-band-centered features perform best with a classifier trained on 2.5\% of the database.
At a cost of 555 training samples, 2,167 true positive and 5,921 false positive SCF and band-structure calculations (39\% of the database), this model finds 88\% of \gls{tcm} candidates.

\section*{Discussion}
In this work, we have presented a unified framework for constructing machine-learning features based on the electronic structure of molecules and materials, leveraging the symmetry preservation and conceptual clarity of physical approaches.
By formalizing the process of featurization as a multistep algorithm, involving the selection of an electronic model, design and application of a spectral operator, and reduction of the spectrum into compact, invariant features, we were able to rapidly design and apply two sets of features to the problems of polymorph similarity and the discovery of transparent-conducting materials.

We have described and investigated a kinetic-operator-based \gls{sorep} method and applied it successfully to distinguishing structurally similar, but electronically diverse, carbon nanotubes.
Using the \glspl{sorep}, metallic and insulating polymorphs were clearly distinguishable, with particular features paralleling $p_z$-tight-binding densities of states.
Applied to the uniqueness analysis of BaTiO$_3$ structures from the \gls{mc3d} feedstock, these kinetic features also showed a remarkable ability to highlight configurations that are missed by the currently employed atomic-structure based method.
In combination with advanced clustering algorithms, dimensionality reduction schemes, and intrinsic dimension analyses such as those implemented in DADApy, the physical interpretability of \gls{sorep} features may also lead in future work to a better understanding of the important electronic collective variables in similar data sets.

A second \gls{sorep} featurization based on a single-shot evaluation of the Kohn-Sham Hamiltonian was then investigated for the more complex and compositionally diverse problem of transparent-conducting material discovery within the \gls{mc3d}.
By leveraging the \gls{sorep} framework, minimal modifications to the kinetic \glspl{sorep} were identified and remedied, producing features remarkably similar to self-consistent \gls{dft} \gls{dos} features at a fraction of the computational effort.
Used to train a random forest classifier, these features allowed for the ``discovery'' of 76\% of materials in the \gls{mc3d} which meet common TCM screening criteria while relying on only 1\% of the database for reference data.
The success of this approach is not only due to their inherently physical information content but also to their interpretability, allowing researchers to select the most meaningful features by leveraging their scientific knowledge and experience.
The strong data efficiency of \gls{sorep} features, i.e. that they can be used to train accurate models with little training data, opens the door to their application in learning difficult-to-compute properties or predictions from levels of theory beyond \gls{dft} which are much more computationally expensive.

\section*{Acknowledgements}
A.M. acknowledges support from the ICSC -- Centro Nazionale di Ricerca in High Performance Computing, Big Data and Quantum Computing, funded by European Union - NextGenerationEU -- PNRR, Missione 4 Componente 2 Investimento 1.4.
N.M. acknowledges support from the NCCR MARVEL, a National Centre for Competence in Research, funded by the Swiss National Science Foundation (project number 51NF40-205602).
A.Z. acknowledges M. Kotiuga and M. Uhrin for fruitful and helpful discussions.

\bibliography{references}

\end{document}